# First-Principles Study of Novel Lead-Free Double Perovskite $\beta_2SnGeX_6$ (β = K, Rb; X = Cl, Br, I) for thermomechanical, optoelectronic and outstanding thermoelectric applications

Jubair Hossan Abir[1], Tauhidur Rahman[1], S.S.B. Pallab[2], Md. Sharear Aman[1],
Raihana Shams Islam[1*], Saleh Hasan Naqib[1*]

[1]Department of Physics, University of Rajshahi, Rajshahi 6205
[2]Department of Electrical and Electronic Engineering, University of Rajshahi, Rajshahi 6205
*Corresponding authors; Emails: *salehnaqib@yahoo.com, *rsislam@ru.ac.bd

**Abstract**

In this study, the structural, mechanical, electronic, optical, and thermoelectric properties of the novel lead-free halide double perovskite series $\beta_2SnGeX_6$ (β = K, Rb; X = Cl, Br, I) are systematically investigated using density functional theory (DFT). Structural optimizations are performed using the PBE-GGA, while accurate electronic and transport properties are evaluated using the Tran-Blaha modified Becke-Johnson (TB-mBJ) potential via the FP-LAPW method in the WIEN2k code. Semiclassical Boltzmann transport equations are solved via the BoltzTrap package to evaluate thermoelectric performance. Calculated formation energies, Goldschmidt's tolerance factors, and octahedral factors confirm that all six compounds exhibit robust thermodynamic stability within a highly symmetric cubic geometry [*Fm*–3*m* (no. 225) space group]. Mechanical analysis derived from elastic parameters characterizes the entire series as fundamentally ductile, ensuring high processing elasticity and resistance to micro-cracking during device manufacturing. Electronic band structures reveal direct bandgaps located precisely at the Γ-point, showing exceptional composition-dependent tunability from 1.44 eV down to 0.64 eV via progressive halogen substitution (Cl>Br>I). The wide gap chloride variations are optimized for single-junction photovoltaic absorbers, while the narrower-gap bromide and iodide analogs show immense promise for tandem solar architectures and near-infrared photodetectors. Thermoelectrically, heavy constituent atoms introduce strong lattice anharmonicity and intense high-temperature Umklapp phonon scattering, significantly suppressing lattice thermal conductivity. Combined with low carrier effective masses that optimize electrical transport, the iodide compounds achieve higher power factors and outstanding dimensionless figures of merit ($ZT \approx 2.4$ for $K_2SnGeI_6$ at 1000 K). Ultimately, the lead-free β2SnGeX$_6$ family emerges as an environmentally benign and versatile platform for next-generation green optoelectronics and solid-state waste-heat recovery.

# 1 Introduction

Conventional power generation systems dependent on fossil fuels are gradually losing favor among the scientific community because of the rapid depletion of global fuel reserves and the serious environmental pollution caused by their use. This shortage of energy resources has encouraged researchers to explore alternative methods, including the conversion of waste heat into electrical energy [1]. In this scenario, non-traditional power generation technologies such as solar cell fabrication, photonic sensor development, thermoelectric generators, and luminescent material enhancement have attracted significant attention as alternatives to fossil fuel-based systems to fulfil the future energy demands without damaging the environment [2]. Due to its abundant availability and environmental friendliness, solar energy is regarded as one of the most favorable options among renewable energy sources such as solar and wind power [3–5]. Solar cells represent one of the most effective technologies to exploit solar energy which directly transform sunlight into electricity via the photovoltaic effect providing a sustainable and environmentally friendly power solution [6–10]. First-generation solar cells, primarily based on silicon, are the most extensively utilized due to their wide availability, environmental friendliness, and high stability. [11]. In terms of efficiency, monocrystalline silicon solar cells have demonstrated a laboratory conversion efficiency around 26%. [12]. However, their practical use is limited by complicated fabrication methods and high manufacturing costs, prompting researchers to find alternative materials for solar energy conversion [13]. As a result, researchers are increasingly interested in developing photovoltaic (PV) materials that are affordable, easy to make, and environmentally friendly [14]. Thin-film solar cells are considered second-generation photovoltaic (PV) technology because they are cheaper to produce although their efficiency is relatively lower [15,16]. In this context, perovskite solar cells have gained considerable attention as third-generation photovoltaic technology because they are widely available, inexpensive, and more efficient compared with silicon-based solar cells [17,18]. Double perovskite materials are promising candidates for renewable and clean energy applications because they have unique & excellent optoelectronic and thermoelectric properties [19]. These properties primarily arise from their distinctive crystal structure, especially the highly flexible corner-sharing $BX_6$ octahedral network [20]. This structure allows materials to offer several useful features, including distinct and adjustable band gaps, strong absorption coefficients of visible light, small carrier effective masses, better carrier mobility, elevated charge diffusion, and high thermoelectric performance [21,22]. In addition to solar cells and thermoelectric generators, perovskite materials are also used in other devices like light emitting diodes (LEDs), ultraviolet detectors, ultraviolet sensors, lasers, and energy conversion systems [23–27]. The efficiency of perovskite solar cells has increased greatly, rising from 3.8% in 2009 to more than 24.2% in 2019, which has strengthened their position as one of the most promising photovoltaic technologies [28]. Organic-inorganic hybrid perovskites generally have the formula $ABX_3$ [29]. In this structure, A is a relatively cation, usually from the s/p block elements, B is a metal cation predominantly from d/f block elements, and X denotes an anion, typically halogen or oxygen [30]. During the past decade, several attempts have targeted improved photoelectrochemical performance via the synthesis of organic-inorganic hybrid composites or solids that combine an organic molecule with a lead halogen perovskite [31–34]. Lead-based halide perovskites, such as $CsPbBr_3$, have become exceptional semiconductors for diverse optoelectronic applications, especially in photovoltaics [35]. However, the performance of Pb-based halide perovskites is strongly affected by moisture, temperature, and other environmental factors, which is its major limitation. Another demerit is the toxicity of lead [36]. Moreover, halide vacancy migration reduces the stability of these materials [37]. As a result, many experimental and theoretical studies have focused on addressing this problem by replacing lead with suitable elements that are environmentally safe, stable at room temperature, and suitable for long-term efficient production. Replacing toxic $Pb^{2+}$ions in perovskites with divalent metal ions such as tin ($Sn^{+2}$) and germanium ($Ge^{+2}$) does not significantly distort the perovskite structure, because the outer shell of Sn or Ge is almost similar to that of Pb [38]. Such substitution either partially or fully, reduces toxicity without compromising the efficiency [39–42]. However, recent research has increasingly focused on halide double perovskites (HDPs), which offer a promising route to stable and lead-free photovoltaic materials. In these compounds, two $Pb^{2+}$ cations are replaced by one monovalent (+1) and one trivalent (+3) non-toxic metal cation, forming the general formula $A_2BB'X_6$ ,which helps improve stability and reduce toxicity [43]. In double perovskite, substituting half of the B cations with suitable B′-elements forms the $A_2BB'X_6$ structure. This replacement induces significant and beneficial properties due to the ordered arrangement of B and B′ cations within the structure [44]. This indicates that double perovskite has the ability to substitute the organic–inorganic perovskite materials and provide a more stable alternative [2]. Mechanically and dynamically stable lead-free halide double perovskites are considered important in this field. To identify promising candidates, several experimental and computational studies have been carried out. The mechanically and dynamically stable lead-free halide double perovskites are considered important in this field. To identify promising candidates, several experimental synthesis and computational studies have been carried out [45–47]. However, these studies generally report two main limitations: (i) either the bandgap is indirect or (i) direct, but with larger values that are not suitable for solar cell applications. Several synthesized and characterized materials have shown good environmental friendliness and stability. Few examples are $Cs_2AgInX_6$ [47], $Cs_2AgSbBr_6$ [48], $Cs_2AgBiX_6$ [49,50] and $Cs_2CuInX_6$ [51], where X (Cl, Br, I). However, most of these materials exhibit indirect band gaps, limiting their suitability for photovoltaic applications [47,48,52,53]. Bi-based chloride and bromide double perovskites generally show wide band gaps, which limits their photovoltaic potential. Although replacing Bi with In produced a direct band gap of 3.84 eV in a bromide double perovskite [54], this value remains too large for efficient solar cell applications. The chloride and bromide of the Bi-type DPs were observed to have wide bandgaps, placing them at a disadvantage for photovoltaic applications. Recently, $Rb_2TlB'I_6$ (B′ = As, Ga) double perovskites have been reported to exhibit good structural and thermodynamic stability. $Rb_2TlGaI_6$ possesses a 1.2 eV indirect band gap, while $Rb_2TlAsI_6$ shows a 1.09 eV direct band gap with strong visible-light absorption, making them promising for solar energy applications [55]. Similarly, the optoelectronic and transport properties of $Na_2AuMX_6$ (M = Al, Ga; X = Cl, Br, I) double perovskites suggest that $Na_2AuAlBr_6$ and $Na_2AuGaBr_6$ are promising for solar cells applications because of their strong visible-light absorption, while $Na_2AuAlI_6$ and $Na_2AuGaI_6$ are more suitable for infrared optoelectronic devices [56]. The first study on Sn based halide perovskites reported $Cs_2SnI_6$ as a hole transport material for dye-sensitized solar cells (DSSCs) [57]. The compound crystallizes in a cubic *Fm*–3*m* (#225) type structure at ambient temperature, and exhibits improved air/moisture stability, n-type electrical conductivity, a direct band gap (1.3 eV), and strong visible light absorption, making it highly suitable for photovoltaic applications [57–59]. Substituting 'Cs' with the smaller 'Rb' cation forms $Rb_2SnI_6$, leading to significant changes in crystal structure and the corresponding optoelectronic properties [60]. $Rb_2SnI_6$ exists in three different crystallographic phases (cubic, tetragonal, and monoclinic) [60,61] and, similar to $Cs_2SnI_6$, shows n-type electrical conductivity. Similarly, several other Sn-based compounds from the $A_2BX_6$ family have also been reported, such as $K_2SnX_6$ (X = Cl, Br, I) [62], $Cs_2Sn(I, Br)_2$ [63] and $Cs_2SnX_6$ (X = Cl, Br, I) [64,65]. During the past few years, research on $A_2BX_6$ based perovskites has broadened to include other tetravalent cations, particularly Ge, opening a promising pathway for substituting Sn in $A_2SnX_6$. In this context, Ge was introduced into $Cs_2SnI_6$ to investigate its influence on the structural, electronic, elastic, and thermodynamic properties of $Cs_2Sn_{(1-x)}Ge_{(x)}I_6$ [66]. Similarly, the Sn-Ge based perovskite $Cs_2SnGeX_6$ (where, X = I, Br, Cl), exhibit widely tunable bandgap of 0.46 to 1.45 eV [66].

Following these earlier leads, in this study, the $\beta_2SnGeX_6$ (β = K, Rb; X = Cl, Br, I) compounds were considered and their mechanical stability was examined. The electronic properties were analyzed through band structures and electronic density of states. Furthermore, the optical behavior was investigated using dielectric constants and other related optical parameters, including refractive index and absorption coefficient. In addition, the thermoelectric performance was evaluated using the Boltz-Trap code by calculating temperature-, chemical potential-, and carrier concentration-dependent transport properties along with the thermoelectric figure of merit for potential power generation applications. We have found that the selected compound holds excellent promise as energy harvesting systems. Moreover, this compound possesses attractive mechanical and optoelectronic features suitable for device applications in diverse sectors.

# 2 Computational method

The structural, electronic, mechanical, optical, and thermoelectric properties of the halide-perovskites $\beta_2SnGeX_6$ (β = K, Rb; X = Cl, Br, I) were investigated employing density functional theory (DFT) framework using the WIEN2k package [67,68], which implements the full-potential linearized augmented plane wave (FP-LAPW) method [69]. The Perdew–Burke–Ernzerhof (PBE) functional within the Generalized Gradient Approximation (GGA) was employed at the initial stage to determine the ground-state structural properties [70]. The structural optimization was performed using the FP-LAPW framework, along with the third-order Birch–Murnaghan equation of state [71]. The plane-wave cutoff was set to $RMT \times k_{max} = 8.0$, where RMT and $k_{max}$ represent the smallest muffin-tin radius and the maximum reciprocal-lattice vector in the plane-wave expansion. The first Brillouin zone (BZ) was sampled with a dense k-point mesh of 10000 k-points to obtain accurate potential and charge density and guarantee computational accuracy. In SCF (self-consistent field) calculations, the convergence thresholds for charge and total energy are set at 0.001e and $10^{-4}$, respectively. To obtain more accurate band gaps in these semiconductors, the Trans-Blaha modified Becke-Johnson (TB-mBJ) [72] potential is used. The potential is expressed as: $V_{(x,\sigma)}^{\mathrm{mBJ}}(r) = cV_{(x,\sigma)}^{\mathrm{BR}}(r) + (3c-2)\frac{1}{\pi}\frac{\sqrt{5}}{12}\frac{\sqrt{2t_\sigma(r)}}{\rho_\sigma(r)}$ where, c, $\rho_\sigma(r)$ and $t_\sigma(r)$ denotes the charge convergence factor, the density of states and the kinetic energy, respectively. The c factor can be calculated using the following expression: $c = \alpha + \left(\beta\frac{1}{V_{cell}}\int_{cell} d^3r\,\frac{|\nabla\rho(r)|}{\rho(r)}\right)^{1/2}$, where α and β are constants, whose values are adjusted in Wien2K to obtain the accurate electronic band structure. TB-mBJ was used to calculate the electronic, optoelectronic and thermoelectric properties of the materials under investigation. A 21 × 21 × 21 k-point mesh was employed TB-mBJ based optical properties calculations. The crystal structures were generated and visualized using VESTA software [73]. The thermoelectric transport properties were evaluated through solution of the semi-classical Boltzmann transport equations using the BoltzTraP package [74] within the framework of Boltzmann transport theory. In this approach, the electrical conductivity($\sigma$), Seebeck coefficient($S$), and electronic thermal conductivity ($\kappa_e$) are obtained from the energy-dependent transport distribution function. The transport coefficients are expressed as follows:

$$\sigma_{\alpha\beta}(T,\mu) = \frac{1}{V}\int \Sigma_{\alpha\beta}(E)\left[-\frac{\partial f_\mu(T,E)}{\partial E}\right] dE \;\; 1 \tag{1}$$

$$S_{\alpha\beta}(T,\mu) = \frac{1}{eTV\sigma_{\alpha\beta}}\int \Sigma_{\alpha\beta}(E)(E-\mu)\left[-\frac{\partial f_\mu(T,E)}{\partial E}\right] dE \tag{2}$$

$$\kappa_{\alpha\beta}^{e}(T,\mu) = \frac{1}{e^2TV}\int \Sigma_{\alpha\beta}(E)(E-\mu)^2\left[-\frac{\partial f_\mu(T,E)}{\partial E}\right] dE \tag{3}$$

Where $E$ denotes the carrier energy, μ is the chemical potential, $f_\mu$ represents the Fermi-Dirac distribution function, V is the unit-cell volume and $\alpha$ and $\beta$ indicate the Cartesian directions. The energy projected conductivity tensors can be expressed as:

$$\Sigma_{\alpha\beta}(E) = \frac{e^2}{N} \sum_{i,k} \tau_{i,k} v_\alpha(i,k) v_\beta(i,k) \delta(E - E_{i,k}) \tag{4}$$

where $N$ represents the number of k-points, $i$ is the band index, $v$ represents the carrier group velocity, and $\tau$ is the relaxation time. The transport properties were analyzed using the constant relaxation time approximation, where the electrical conductivity and electronic thermal conductivity depend on $\tau$, while the Seebeck coefficient remains independent of the relaxation time.

# 3 Results and discussion

## 3.1 Structural parameters and phase stability

The polyhedral view of $\beta_2SnGeX_6$ (β = K, Rb; X = Cl, Br, I) double perovskite structure is presented in **Fig. 1.** The double perovskite structure crystalizes in the cubic structure with space group *Fm*-3*m* (#225). The unit cell contains four formula units consisting of 40 atoms in total. In this crystal structure, the Sn atom occupies the (0, 0, 0) Wyckoff position, while the Ge atom is located at (0.5, 0.5, 0.5). The A-site alkali cation (K or Rb) resides at the (0.25, 0.25, 0.25) position, and the halogen atoms occupy the (x, 0, 0) sites. The optimized internal structural parameter x for the halogen site was obtained as 0.25655, 0.25627, and 0.25561 for $K_2SnGeCl_6$, $K_2SnGeBr_6$, and $K_2SnGeI_6$, respectively, while the corresponding values for the Rb-based compounds are 0.25661, 0.25616, and 0.25577.

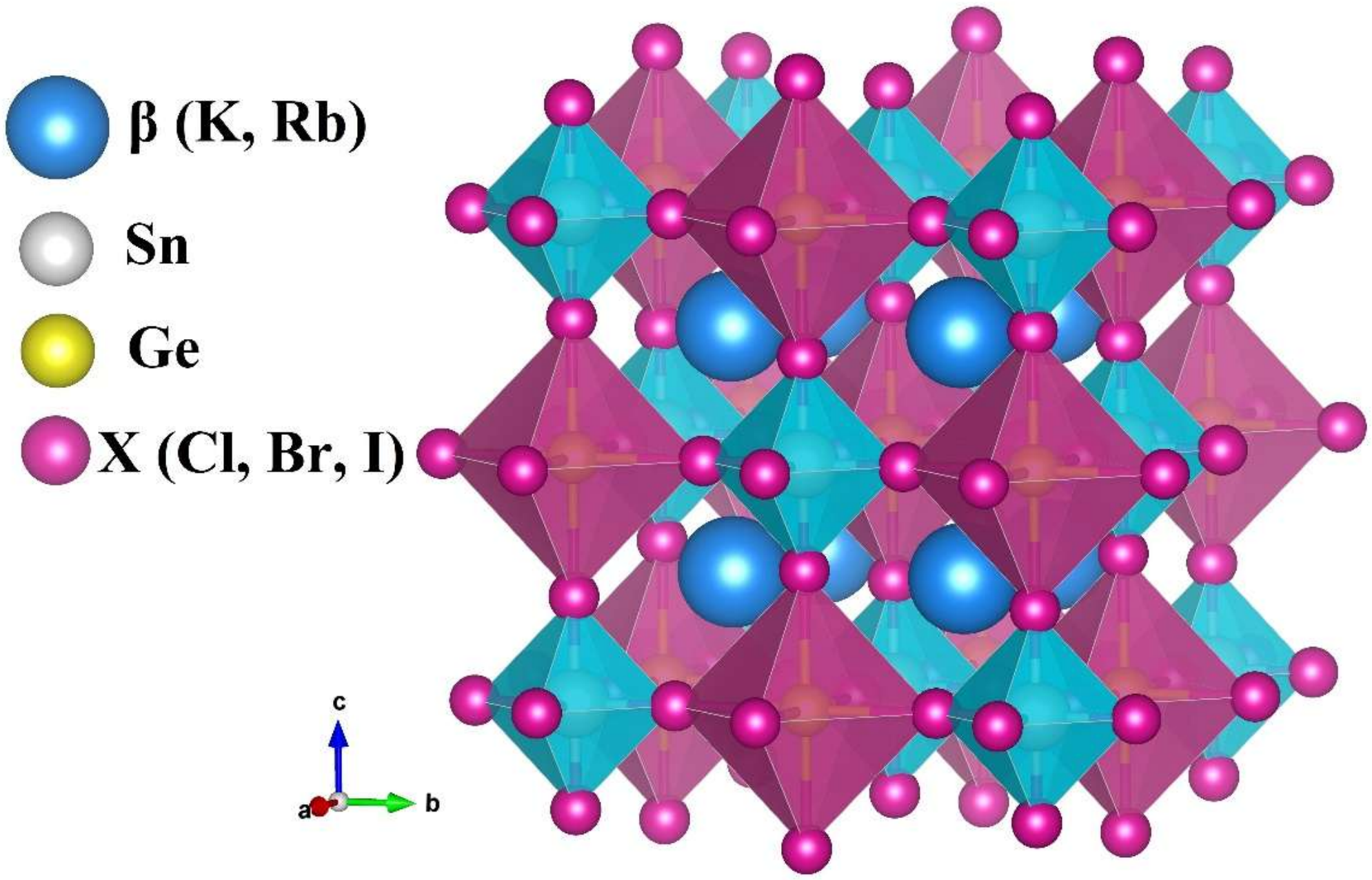

**Fig. 1.** The Crystal Structure of $\beta_2SnGeX_6$ (β = K, Rb; X = Cl, Br, I).

**Table 1.** Calculated values of lattice constant ($a_0$), tolerance factor ($t_f$), octahedral factor ($\mu$), and energy of formation (E_F).

| Compound | $a_0$ (Å) | $t_f$ | $\mu$ | E_F (eV/atom) |
|---|---|---|---|---|
| $K_2SnGeCl_6$ | 10.840 | 0.876 | 0.539 | -5.764 |
| $K_2SnGeBr_6$ | 11.361 | 0.867 | 0.497 | -5.285 |
| $K_2SnGeI_6$ | 12.167 | 0.855 | 0.443 | -4.615 |
| $Rb_2SnGeCl_6$ | 10.876 | 0.896 | 0.539 | -5.765 |
| $Rb_2SnGeBr_6$ | 11.407 | 0.887 | 0.497 | -5.288 |
| $Rb_2SnGeI_6$ | 12.196 | 0.873 | 0.443 | -4.619 |

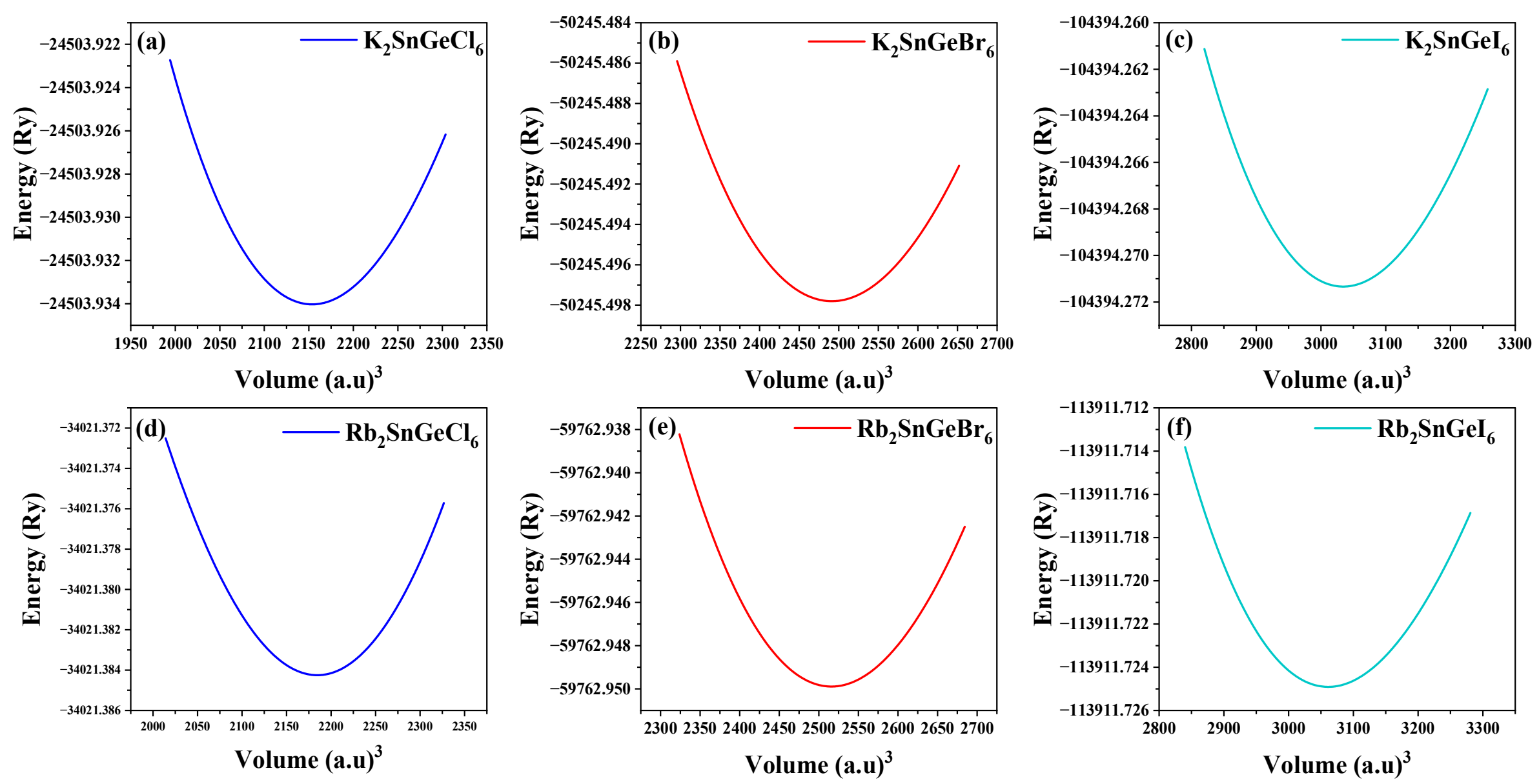

**Fig. 2.** Ground state energy as a function of volume for $\beta_2SnGeX_6$ (β = K, Rb; X = Cl, Br, I).

Fig. 2 illustrates the ground-state energy (the lowest energy state at absolute zero temperature and zero pressure) as a function of the volume for $\beta_2SnGeX_6$ (β = K, Rb; X = Cl, Br, I). The equilibrium structure is determined through total-energy minimization using the Birch–Murnaghan equation of state (EOS) [71]. The optimized lattice parameters obtained from this fitting procedure are summarized in **Table 1.**

In this ordered double perovskite framework, both Sn and Ge cations are octahedrally coordinated by six halogen atoms, forming $SnX_6$ and $GeX_6$ octahedral units. These octahedra are arranged alternately and interconnected through corner-sharing networks that from a linear Sn-X-Ge linkage with a bond angle of 180°. Such a highly symmetric arrangement is a characteristic feature of cubic double perovskites and plays a critical role in maintaining structural rigidity and long-range crystalline order.

**Table 1** summarizes the optimized lattice parameters, tolerance factors, octahedral factors, and formation energies of the studied compounds. A gradual increase in the lattice constant is observed with successive halogen substitution from Cl to Br to I in both alkali-metal series, which can be attributed to the progressive increase in the ionic radius of the halogen anions. Furthermore, the lattice parameter is slightly larger for the Rb-based compounds compared with their K-based counterparts due to the larger ionic radius of the $Rb^+$ cation compared with $K^+$.

To further assess the structural stability of these compounds, the Goldschmidt tolerance factor ($t_f$) and the octahedral factor ($\mu$) were evaluated using the following relations:

$$t_f = \frac{r_A + r_X}{\sqrt{2}(r_B + r_X)} \tag{5}$$

$$\mu = \frac{r_B}{r_X} \tag{6}$$

where $r_A$, $r_B$, and $r_X$represent the ionic radii of the A-site cation, the effective B-site cation, and the halogen anion, respectively. According to the established structural criteria for perovskite systems, cubic stability is generally favored when the tolerance factor lies within the range $0.814 \leq t_f \leq 1.0$, while the octahedral factor falls within $0.416 \leq \mu \leq 0.896$ [75–77]. For the respective atoms in $\beta_2SnGeX_6$ the Shannon's radii [78] are used to calculate the $t_f$ and $\mu$ values. The calculated tolerance factors lie between 0.855 and 0.896, whereas the octahedral factors vary from 0.443 to 0.539. These values are consistent with empirically established stability windows commonly used for perovskite and perovskite-derived frameworks, indicating that all investigated compounds are geometrically compatible with the cubic structure.

Furthermore, the formation energy ($E_{_F}$) of the compounds studied was evaluated to confirm their thermodynamic stability using **Equation (7)**, as follows:

$$E_{_F} = \frac{E_{total}(\beta_2\mathrm{SnGeX}_6) - [2E(\mathrm{A}) + E(\mathrm{Sn}) + E(\mathrm{Ge}) + 6E(\mathrm{X})]}{10} \tag{7}$$

where $E_{\mathrm{total}}$ ($\beta_2SnGeX_6$) is the total energy of the $\beta_2SnGeX_6$ compound, and $E(\beta)$, $E(Sn)$, $E(Ge)$ and $E(X)$ are the total energy of the individual atoms in the crystalline form, including β (where β = K and Rb) and X (where X = Cl, Br and I), respectively. The calculated formation energies are negative for all the compounds, showing that their formation is energetically favorable relative to the constituent elemental reference states. Among the investigated compounds, $Rb_2SnGeBr_6$ exhibits the lowest formation energy which suggests enhanced thermodynamic stability compared to the other compounds.

The combined structural optimization, geometrical factor analysis, and formation energy evaluation consistently demonstrate that the $\beta_2SnGeX_6$ (β = K, Rb; X = Cl, Br, I) compounds are both structurally and thermodynamically stable in the cubic phase. The halogen dependent lattice expansion, the favorable tolerance and octahedral factors further indicate the compositional tuning through A-site and X-site substitution provides an effective route for controlling the crystal chemistry of this lead-free double-perovskite family.

## 3.2 Mechanical and Elastic Properties

The elastic constants of a solid are fundamental parameters that provide insight into its mechanical stability, stiffness, and structural response to external stress. For a cubic crystal system, there are only three independent elastic constants: $C_{11}$, $C_{12}$ and $C_{44}$. The calculated elastic constants for the $\beta_2SnGeX_6$ (β = K, Rb; X = Cl, Br, I) compounds are presented in **Table 2**. To evaluate the mechanical stability of these compounds, we apply the generalized Born stability criteria for cubic crystals [79]: $(C_{11}-C_{12}) > 0$, $C_{11} > 0$, $C_{44} > 0$, and $(C_{11} + 2C_{12}) > 0$. As shown in **Table 2**, all six investigated compounds satisfy these criteria, confirming that they are mechanically stable in their respective crystalline phases. The elastic constants $C_{11}$, $C_{12}$, and $C_{44}$ are among the most important components of the elastic stiffness tensor. These parameters describe how a crystal resists different types of deformation and provide insight into interatomic bonding, mechanical stability, anisotropy, and lattice dynamics. $C_{11}$ gives the resistance to uniaxial compression or tension along a principal crystallographic axis, $C_{12}$ measures the coupling between strains in two perpendicular directions, and $C_{44}$ determines the resistance against shear. All the elastic constants are low, implying that the compounds under investigation are relatively soft solids.

The macroscopic mechanical behavior of the $\beta_2SnGeX_6$ compounds was studied through the bulk modulus (B), shear modulus (G), and Young's modulus (Y). The bulk modulus (B) describes material's resistance to volume change under hydrostatic pressure, whereas the shear modulus (G) measures resistance to reversible shear deformation. Young's modulus (Y) represents the overall stiffness of the material. A clear trend is observable upon halogen substitution as the atomic size of the halogen increases from Cl to I. The values of B, G, and Y monotonically decrease for both the K and Rb based series. For example, in the $K_2SnGeX_6$ series, Y decreases from 26.996 GPa for X = Cl to 19.192 GPa for X = I. This reduction in elastic moduli indicates that the compounds become softer and more compressible with heavier halogens. This can be attributed to the increased atomic radii and corresponding larger bond lengths which leads to weaker interatomic bonding forces.

**Table 2.** Calculated elastic constants $C_{ij}$ (GPa), bulk modulus B (GPa), Cauchy pressure $C_{cauchy}$ (GPa), shear modulus G (GPa), Young's modulus Y (GPa), Poisson's ratio ν, Pugh's ratio B/G, Zener anisotropy factor (A)**,** Machinability index $\mu^M$, Grüneisen parameter γ and Melting temperature of $\beta_2SnGeX_6$ (β = K, Rb; X = Cl, Br, I).

| Compound | $C_{11}$ | $C_{12}$ | $C_{44}$ | $C_{Cauchy}$ | B(GPa) | G(GPa) | Y(GPa) | B/G | $v$ | A | $\mu^M$ | $\gamma$ | $T_m$ (K) |
|---|---|---|---|---|---|---|---|---|---|---|---|---|---|
| $K_2SnGeCl_6$ | 54.01 | 9.47 | 5.75 | 3.72 | 24.32 | 10.27 | 26.10 | 2.37 | 0.315 | 0.258 | 4.2 | 1.87 | 872.21 |
| $K_2SnGeBr_6$ | 49.30 | 8.60 | 5.74 | 2.86 | 22.17 | 9.82 | 25.66 | 2.26 | 0.307 | 0.282 | 3.9 | 1.82 | 844.36 |
| $K_2SnGeI_6$ | 38.51 | 5.24 | 3.96 | 1.28 | 16.33 | 7.36 | 19.19 | 2.22 | 0.304 | 0.238 | 4.1 | 1.80 | 780.57 |
| $Rb_2SnGeCl_6$ | 53.79 | 10.01 | 6.83 | 3.18 | 24.60 | 11.14 | 29.03 | 2.21 | 0.303 | 0.312 | 3.6 | 1.79 | 870.88 |
| $Rb_2SnGeBr_6$ | 46.99 | 8.04 | 5.89 | 2.15 | 21.02 | 9.75 | 25.32 | 2.16 | 0.299 | 0.302 | 3.6 | 1.77 | 830.70 |
| $Rb_2SnGeI_6$ | 37.10 | 5.01 | 4.58 | 0.42 | 15.70 | 7.79 | 20.06 | 2.02 | 0.287 | 0.285 | 3.4 | 1.69 | 772.26 |

The ductile or brittle nature of the materials is a critical factor for their potential industrial and engineering applications. This behavior was evaluated using Pugh's ratio (B/G), where a value greater than 1.75 indicates ductile behavior, while a value below 1.75 suggests brittleness [80]. For all the $\beta_2SnGeX_6$ compounds, the calculated B/G ratios range from 2.015 to 2.369 that clearly exceed the critical threshold of 1.75. This indicates that all these compounds are inherently ductile. The ductility is further corroborated by Poisson's ratio (ν). According to Frantsevich's rule [81], a material is ductile if $\nu > 0.26$. For the studied compounds, the calculated values of ν range from 0.287 to 0.315 which exceeds this threshold reaffirming their ductile nature.

To further investigate the bonding characteristics, we analyzed the Cauchy pressure ($C$), defined as ($C_{12}$-$C_{44}$). According to Pettifor's criteria [82], a positive Cauchy pressure is indicative of significant metallic bonding and a ductile nature, whereas a negative value points to directional covalent bonding and brittleness. As detailed in **Table 2**, the Cauchy pressures for all compounds are positive, ranging from 0.422 to 3.723 GPa, which is in excellent agreement with the conclusions drawn from Pugh's and Poisson's ratios.

Finally, the elastic anisotropy of a material is important for understanding the probability of microcracks developing in the material. The Zener anisotropy factor (A) for cubic crystals was calculated as $A = 2C_{44}/(C_{11}-C_{12})$ [83]. For a perfectly isotropic solid (A = 1), any deviation from unity signifies elastic anisotropy. The calculated A values for the $\beta_2SnGeX_6$ compounds lie between 0.238 and 0.312. Since these values deviate significantly from 1, it demonstrates that these compounds possess strong elastic anisotropy which means their mechanical responses are highly dependent on the crystallographic direction. The machinability index ($\mu^M$) ranges from 3.4 to 4.2, suggesting a very high level of structural workability for device fabrication [84-87]. All these compounds are expected to exhibit very high level of dry lubricity as well.

The degree of lattice anharmonicity was evaluated through the calculation of the Grüneisen parameter (γ), which describes the volume dependence of phonon frequencies and the strength of phonon-phonon scattering processes [80]. The calculated γ values for the $\beta_2SnGeX_6$ (β = K, Rb; X = Cl, Br, I) compounds are listed in **Table 2**. The γ values vary from 1.69 to 1.87 that indicates moderate lattice anharmonicity in these materials. For both the K- and Rb-based systems, γ decreases gradually with the substitution of heavier halogens from Cl → Br → I. This behavior suggests a reduction in anharmonic lattice vibrations with increasing halogen atomic mass, which can influence phonon scattering mechanisms and consequently affect the lattice thermal conductivity. Moreover, the estimated melting temperatures confirm the moderate thermal endurance of the studied compounds, which is an important requirement for stable optoelectronic device performance.

The lattice thermal conductivity, $\kappa_{ph}$, is a key parameter for evaluating the thermoelectric properties and to determine the energy-conversion efficiency of thermoelectric materials under the high temperature operating conditions. A reduced $\kappa_{ph}$ is highly desirable because it helps to maintain a large temperature gradient, thereby improving overall thermoelectric efficiency. The suppression of lattice thermal conductivity is mainly governed by phonon-scattering mechanisms which play a central role in limiting heat propagation in crystalline solids [84]. In this work, the temperature dependent lattice thermal conductivity of $\beta_2SnGeX_6$ compounds was evaluated using Slack's empirical formula [85,86].

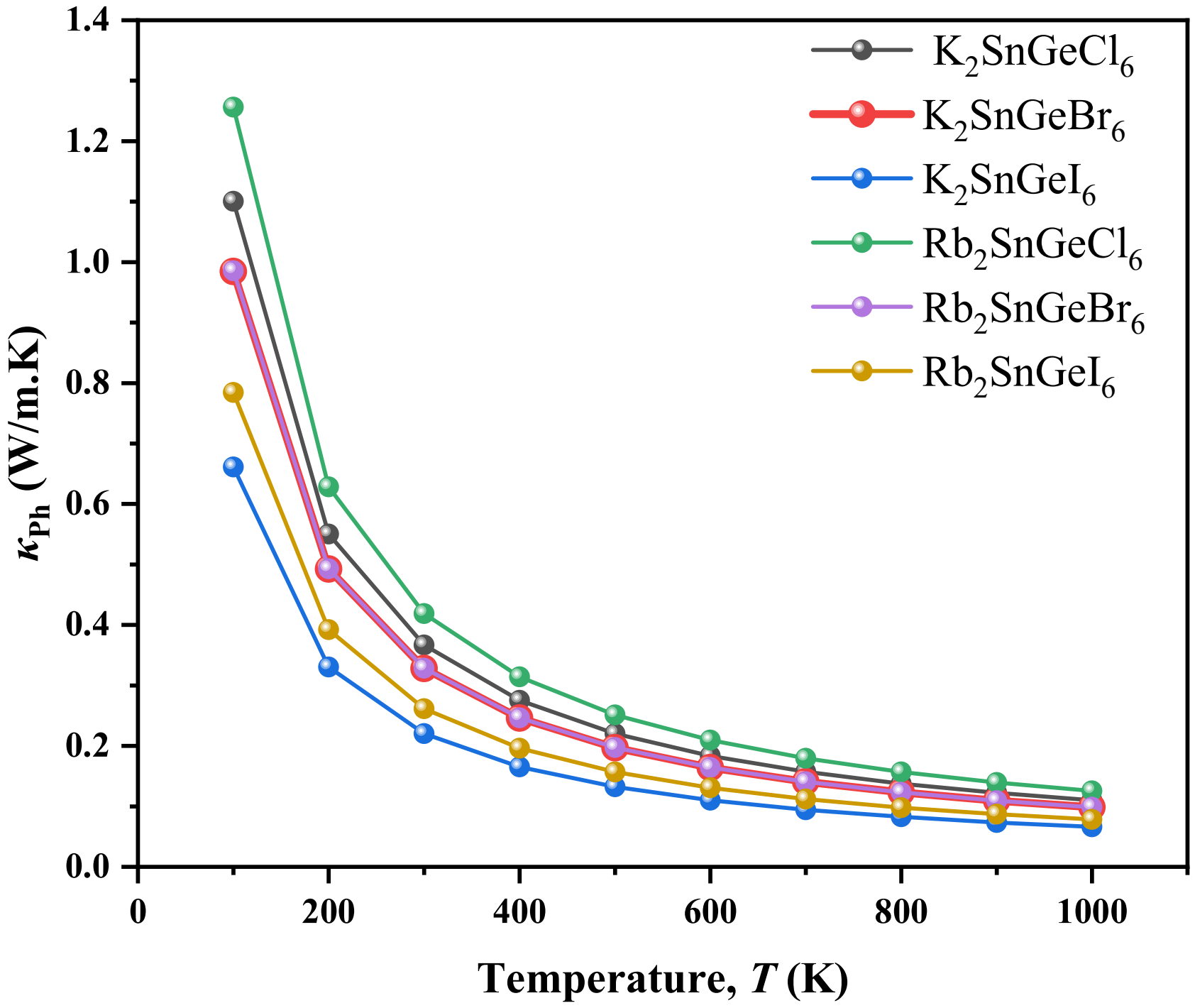

**Fig. 3.** Lattice thermal conductivity of $\beta_2SnGeX_6$ (β = K, Rb; X = Cl, Br, I) as a function of temperature.

The temperature dependence of lattice thermal conductivity for the double perovskites $\beta_2SnGeX_6$ ($\beta$ = K, Rb; X = Cl, Br, I) is presented in

Fig. 3. The calculated ($\kappa_{ph}$) decreases monotonically with increasing temperature for all investigated compounds. This behavior is commonly found in crystalline solids where thermal transport is mainly governed by lattice vibrations. The decrease in $\kappa_{ph}$ with temperature originates mostly from enhanced phonon-phonon scattering, particularly Umklapp Scattering, which becomes more pronounced at elevated temperatures and significantly limits heat transport through the lattice.

A similar pattern is also observed across the halide series, where the lattice thermal conductivity decreases consistently when moving from Cl to Br and further to I. This pattern can be generated due to the increasing atomic mass of the halogen atoms and the subsequent reduction in phonon group velocity which suppresses lattice heat transport. Heavier atoms also introduce stronger lattice anharmonicity, which further contributes to the reduction of thermal conductivity [87–90]. In addition, the substitution of the β-site cation from K to Rb slightly modifies the phonon transport behavior due to differences in ionic size and mass influencing the vibrational properties of the crystal lattice.

Overall, the relatively low lattice thermal conductivity estimated for these compounds indicates that their thermal transport properties are strongly influenced by heavy constituent atoms, complex lattice structures, and anharmonic lattice vibrations. Such characteristics are highly desirable for thermoelectric materials because suppressed lattice heat transport can improve thermoelectric efficiency by maintaining a high temperature gradient across the material [90,91]. It has to be noted that since these compounds are semiconductors the electronic contribution to the total thermal conductivity is expected to be small.

## 3.3 Electronic Properties

### 3.3.1 Band Structure and density of states

Electronic properties play a significant role in determining the application of materials in modern electronic and optoelectronic devices. In particular, double halide perovskites have emerged as a major focus of research due to their tunable band gaps and promising charge transport characteristics. Understanding the electronic behavior of these materials is essential for evaluating their potential in applications such as solar cells, photodetectors, and other semiconductor devices. The electronic properties of the double halide perovskites $\beta_2SnGeX_6$ (β = K, Rb; X = Cl, Br, I) were investigated through band-structure, electronic energy density-of-states (DOS), and carrier effective-mass analyses in order to assess their suitability for electronic and optoelectronic applications. The electronic band structure of $\beta_2SnGeX_6$ (β = K, Rb; X = Cl, Br, I) were calculated along high symmetry paths of the first Brillouin Zone (FBZ) and the resulting band dispersions are presented in **Fig. 4**. The horizontal dotted line represents the Fermi level, $E_F$, which is set at zero energy. These figure shows that the valance band maximum (VBM) and the conduction band minimum (CBM) for all the investigated compounds is located at the Γ point, confirming their direct band-gap semiconducting character. This feature is favorable for light absorption and emission, and therefore, beneficial for photovoltaic and other optoelectronic applications.

To obtain reliable electronic band gaps, calculations were carried out using both the PBE-GGA functional and the Tran-Blaha modified Becke-Johnson (TB-mBJ) potential. Although PBE-GGA is computationally efficient and widely used, it is well known to underestimate the band gaps of semiconductors because of the intrinsic limitations of conventional exchange-correlation functionals [68,70]. Therefore, the TB-mBJ potential was also employed to provide improved band-gap estimates, since it has been shown to provide more accurate band gap predictions for semiconductors and insulators [72].

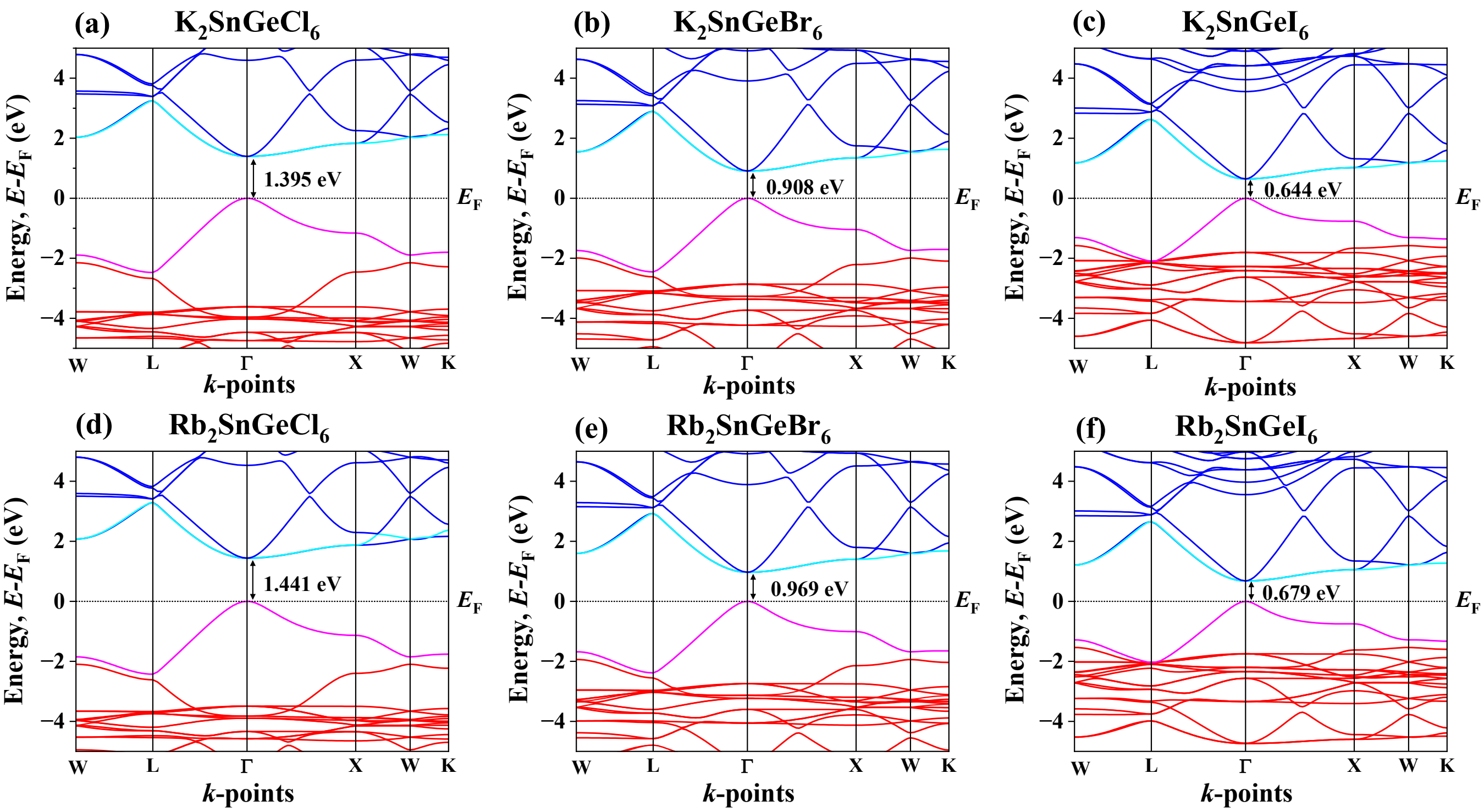

**Fig. 4.** Calculated band structures of (a) $K_2SnGeCl_6$ (b) $K_2SnGeBr_6$ (c) $K_2SnGeI_6$ (d) $Rb_2SnGeCl_6$ (e) $Rb_2SnGeBr_6$ and (f) $Rb_2SnGeI_6$ compounds. All band structures are plotted along the high symmetry path of the FBZ. The Fermi level is set to zero energy (horizontal dot line) for reference.

**Table 3.** Calculated band gaps of $\beta_2SnGeX_6$ ($\beta$ = K, Rb; X = Cl, Br, I).

| Compound | Electronic band gap (eV) [PBE-GGA] | Electronic band gap (eV) [TB-mBJ] |
|---|---|---|
| $K_2SnGeCl_6$ | 0.886 | 1.395 |
| $K_2SnGeBr_6$ | 0.533 | 0.908 |
| $K_2SnGeI_6$ | 0.436 | 0.644 |
| $Rb_2SnGeCl_6$ | 0.926 | 1.441 |
| $Rb_2SnGeBr_6$ | 0.634 | 0.969 |
| $Rb_2SnGeI_6$ | 0.465 | 0.679 |

The calculated band gaps are summarized in **Table 3**. Within the PBE-GGA approximation, the band gaps are 0.886, 0.533, and 0.436 eV for $K_2SnGeCl_6$, $K_2SnGeBr_6$, and $K_2SnGeI_6$, respectively. The corresponding values for the Rb-based compounds are 0.926, 0.634, and 0.465 eV for $Rb_2SnGeCl_6$, $Rb_2SnGeBr_6$, and $Rb_2SnGeI_6$, respectively. When the TB-mBJ potential is applied, the band gaps increase to 1.395, 0.908, and 0.644 eV for $K_2SnGeCl_6$, $K_2SnGeBr_6$, and $K_2SnGeI_6$, respectively, while $Rb_2SnGeCl_6$, $Rb_2SnGeBr_6$, and $Rb_2SnGeI_6$ exhibit band gaps of 1.441, 0.969, and 0.679 eV, respectively. These results confirm that TB-mBJ predicts systematically larger and more realistic band gaps than PBE-GGA, which is consistent with previously reported studies on semiconducting materials [72].

A clear compositional trend is observed across both alkali-metal series: the band gap decreases from Cl to Br to I. This reduction can be attributed to the progressive upward shift of the halogen p-derived valence states as the halogen changes from Cl to Br to I, which narrows the energy separation between the valence and conduction bands [92,93]. In addition, replacement of K by Rb produces a slight increase in band gap for the corresponding halide system. Because the alkali-metal cations contribute minimally to the states near the Fermi level, this behavior is most reasonably associated with structural modification, including changes in lattice parameter, interatomic distance, and orbital overlap [92].

To Further elucidate the electronic structure, the total and partial density of states (TDOS and PDOS) were analyzed to identify the atomic contribution to the electronic band. The upper panels of Fig. 5 represents the total DOS and the lower panels represent corresponding partial DOS for all investigated compounds. The upper valence region is dominated primarily by the halogen p orbitals, whereas the lower conduction region is mainly formed by the s and p states of Sn and Ge. As the halogen atom becomes heavier, the energy of its p orbitals increases. This causes the valence band to shift upward toward the Fermi level, thereby reducing the band gap. This behavior is consistent with earlier theoretical studies on $A_2BB'X_6$-type halide systems [92,93].

The band gap values of 0.644-1.441 eV calculated using TB-mBJ indicate that these materials can be used in different energy related applications. For instance, the Shockley–Queisser limit[94], indicates the ideal band gap of 1.1-1.6 eV for a high performance single-junction solar cell. More specifically, $K_2SnGeCl_6$ and $Rb_2SnGeCl_6$ possess band gaps close to the ideal range for photovoltaic absorber materials. The bromide and iodide based compounds possess narrower band gaps and may be more suitable for visible to near-infrared optoelectronic applications such as photodetectors and absorber layers in tandem architectures [95]. In addition, relatively narrow-gap semiconductors are often of interest for thermoelectric research because enhanced carrier transport can support improved electrical conductivity and thermoelectric response [96]. The iodide compounds $K_2SnGeI_6$ and $Rb_2SnGeI_6$, with band gaps below 0.7 eV are expected to exhibit favorable electrical conductivity and carrier transport properties. The tunability of the band gap through halogen substitution provides an effective strategy for modifying the electronic properties of these materials for specific technological applications.

The effective masses of electrons ($m_e^*$) and holes ($m_h^*$) are important electronic parameters that significantly influence the mobility and separation efficiency of photo-generated charge carriers [97]. These values can be calculated using the following relation [98]:

$$m^* = \hbar^2 \left(\frac{d^2E}{dk^2}\right)^{-1}$$

where $E(k)$ and $k$ represent the eigenvalues of energy band and the wave vector along different directions, respectively, ħ is reduced Planck constant and $m_0$ represents the bare mass of an electron. The calculated effective mass of electron and hole are shown in Table 4.

**Table 4.** Computed effective masses of electrons ($m_e$*) and holes ($m_h$*) close to the zone center of $\beta_2SnGeX_6$ (β = K, Rb; X = Cl, Br, I).

| Compound | $m_e$* | $m_h$* |
|---|---|---|
| $K_2SnGeCl_6$ | 0.23204 | 0.05156 |
| $K_2SnGeBr_6$ | 0.199918 | 0.04087 |
| $K_2SnGeI_6$ | 0.204474 | 0.02397 |
| $Rb_2SnGeCl_6$ | 0.272831 | 0.05316 |
| $Rb_2SnGeBr_6$ | 0.203835 | 0.05123 |
| $Rb_2SnGeI_6$ | 0.209145 | 0.02955 |

It is evident that the smaller alkali-metal cation and larger halogen anion both contribute to reducing the carrier effective mass, thereby enhancing the electronic transport characteristics of these compounds [96,98,99]. In all cases, the effective mass of hole is substantially smaller than the electron effective mass, indicating that hole transport is likely to be more favorable in these compounds. The low effective masses overall also suggest appreciable band dispersion near the band edges, which is beneficial for carrier mobility and electrical transport.

Overall, the present results demonstrate that $K_2SnGeX_6$ and $Rb_2SnGeX_6$ are direct band-gap semiconductors with composition-dependent electronic properties. The band gap can be effectively tuned through halogen substitution, while alkali-metal substitution produces a smaller but noticeable modification through structural effects. Together with the relatively low effective masses, these findings indicate that the studied lead-free double halide perovskites are promising candidates for future photovoltaic, optoelectronic, and thermoelectric applications.

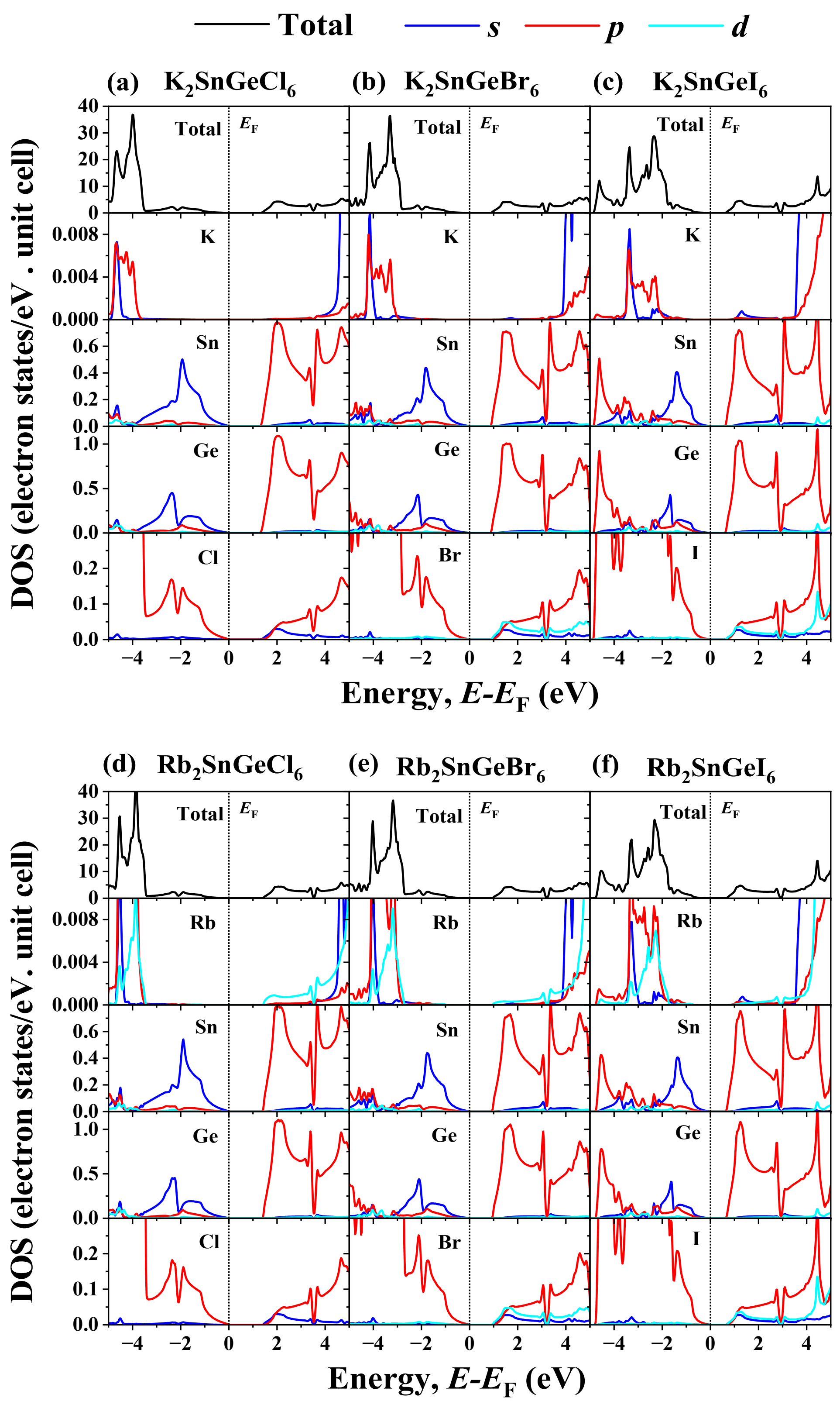

**Fig. 5.** Calculated density of states (DOS) and partial density of states (PDOS) of (a) $K_2SnGeCl_6$ (b) $K_2SnGeBr_6$ (c) $K_2SnGeI_6$ (d) $Rb_2SnGeCl_6$ (e) $Rb_2SnGeBr_6$ and (f) $Rb_2SnGeI_6$ compounds. The Fermi level is set to zero energy (vertical dot line) as a reference.

### 3.3.2 Optical Properties

In optical properties, one studies how a material responds to the irradiation of electromagnetic energy. Optical response was evaluated from frequency-dependent dielectric and related functions governed by the electronic structure up to a photon energy of 14 eV. The static value of dielectric constant, $\varepsilon_1(0)$, defines the static value of refractive index and is necessary to design many optical devices. The estimated static values of dielectric constant, refractive index and reflectivity along [100] polarization directions for $\beta_2SnGeX_6$ (β = K, Rb; X = Cl, Br, I) is shown in **Table 5**. The choice of the alkali metal (A-site) has a negligible impact on the optical response, as the calculated spectra for the potassium and rubidium variants are nearly indistinguishable. This indicates that the optical transitions are predominantly governed by the $SnGeX_6$ octahedra.

**Table 5.** Estimated static values of dielectric constant, refractive index and reflectivity for electric field along [100] polarization direction for $\beta_2SnGeX_6$ (β = K, Rb; X = Cl, Br, I).

| Compound | $\varepsilon_1(0)$ [100] | $n_r(0)$ [100] | $R(0)$ [100] |
|---|---|---|---|
| $K_2SnGeCl_6$ | 3.8167 | 1.9537 | 0.1043 |
| $K_2SnGeBr_6$ | 5.1118 | 2.2610 | 0.1495 |
| $K_2SnGeI_6$ | 7.1765 | 2.6790 | 0.2083 |
| $Rb_2SnGeCl_6$ | 3.7929 | 1.9476 | 0.1034 |
| $Rb_2SnGeBr_6$ | 5.0238 | 2.2414 | 0.1467 |
| $Rb_2SnGeI_6$ | 7.0739 | 2.6598 | 0.2057 |

The macroscopic response is described by the complex dielectric function [**100**], $\varepsilon(\omega) = \varepsilon_1(\omega) + i\varepsilon_2(\omega)$. The frequency-dependent real part, $\varepsilon_1(\omega)$, reflects polarizability and phase velocity, whereas the imaginary part, $\varepsilon_2(\omega)$, describes absorption linked to interband transitions and the band structure [76].

The static limit of the real part, $\varepsilon_1(0)$, increases monotonically with heavier halogen substitution. The calculated $\varepsilon_1(0)$ values are approximately 3.8 for the chloride compounds, 5.0 for the bromides and 7.0 for the iodides [**Fig. 6 (a)**]. The higher value for the iodide compounds is attributed to the larger and more easily polarizable electron clouds of the iodine atoms. At higher photon energies (beyond ~9 eV for iodides), $\varepsilon_1(\omega)$ becomes negative, indicating a transition to metallic-like reflective behavior in the deep ultraviolet region. Large low energy refractive index is desired for efficient wave guides and display devices.

The onset energy of $\varepsilon_2(0)$ perfectly mirrors the fundamental bandgaps of the respective compounds [**101**]. A pronounced redshift toward lower photon energies is observed as the halogen changes from Cl to Br to I [**Fig. 6 (c)**].

Optical absorption starts at the band edge for all compounds, confirming their semiconducting nature at low energies [**Fig. 6 (e)**]. High absorption occurs mainly in the visible and ultraviolet ranges. The absorption spectra exhibit a distinct redshift for the bromide and iodide variants, providing insight into the material's potential application for efficient solar energy conversion.

The zero-frequency reflectivity, R(0), follows the structural trend, starting at roughly 0.10 for the Cl-based compounds, ~ 0.15 for Br, and approaching 0.20 for the I-based compounds [**Fig. 6 (g)**]. Reflectivity increases sharply in the high-energy UV region (>10 eV), coinciding with the negative $\boldsymbol{\varepsilon_1(\omega)}$ domain. Low reflectivity of these compounds implies that they have potential as anti-reflection coating materials.

The static refractive index, n(0), increases from ~ 1.9 for the chlorides to ~ 2.6 for the iodides [**Fig. 6 (b)**]. A high refractive index is advantageous for photovoltaic devices to improve light trapping within the active layer. The extinction coefficient, k($\omega$), shows onset thresholds that map directly to the $\varepsilon_2(\omega)$ and absorption profiles. Optical conductivity starts at the fundamental bandgap and features several sharp peaks in the ultraviolet region, representing strong interband transitions from the valence band to the conduction band. The electron energy loss function, L($\omega$), is used to identify the bulk plasma frequency, indicating where the material transitions from dielectric to metallic behavior [**102**]. The prominent peaks in the loss function are located between 4 eV and 8 eV for the chloride and bromide compounds, representing the high frequency plasmonic resonance energies [**Fig. 6 (f)**].

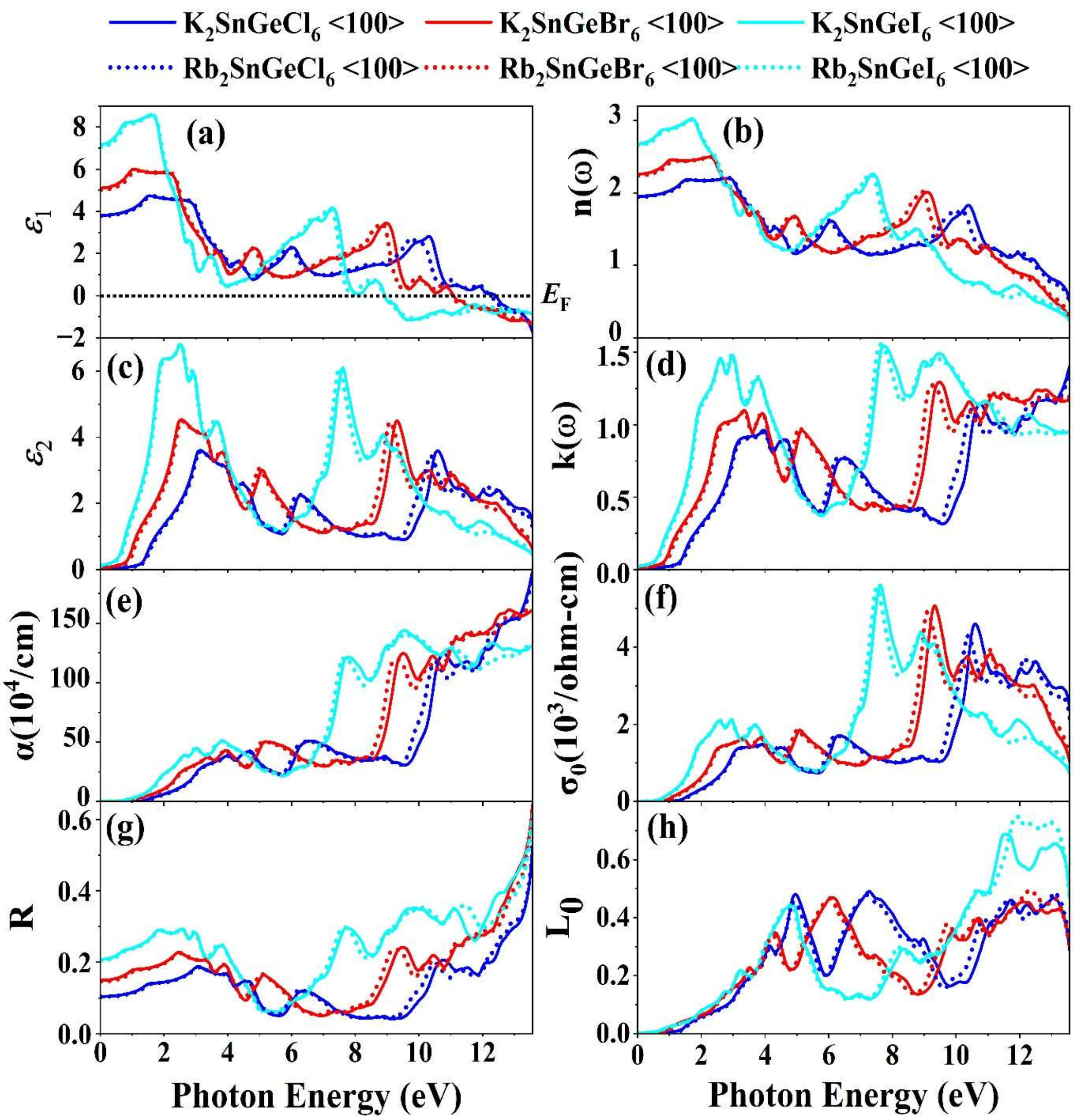

**Fig. 6.** Photon energy dependences of (a) real part, $\varepsilon_1$, (c) imaginary part, $\varepsilon_2$ of dielectric function, (b) refractive index, $n(\omega)$, (d) extinction coefficient, $k(\omega)$, (e) absorption coefficient, α, (f) optical conductivity, $\sigma_o$, (g) loss function, $L_0$ of $\beta_2SnGeX_6$ (β = K, Rb; X = Cl, Br, I) along [100] polarization direction, respectively.

### 3.3.3 Thermoelectric Property

Fig. 7 and Fig. 8 present a systematic theoretical evaluation of the electronic transport properties of two series of halide double perovskites, $\beta_2SnGeX_6$ (β = K, Rb; X = Cl, Br, I), using semi-classical Boltzmann transport theory within the constant relaxation time approximation. In Fig. 7, the transport coefficients are plotted as a function of chemical potential at several fixed temperatures (denoted by red, cyan, and blue curves, corresponding to 300 K, 600 K, and 900 K). Fig. 8 provides the temperature-dependent thermoelectric behavior at constant chemical potential of the materials. The dashed vertical line at 0 eV in Fig. 7 represents the Fermi level. The calculated electronic structures reveal moderate band gaps, which are advantageous for thermoelectric applications because they balance carrier concentration with transport efficiency, indicating that these compounds are promising candidates for efficient thermoelectric applications.

The Seebeck coefficient, representing the voltage generated under a temperature gradient, is shown in Fig. 7(a-f) as a function of chemical potential at different temperatures. The Seebeck coefficient |S| decreases with increasing temperature in the range of ±1 to 0 eV, consistent with the well-known inverse relationship between temperature and thermopower. Doping is simulated by shifts in the chemical potential under the rigid band approximation. Negative chemical potentials ($\mu < 0$) are associated with positive Seebeck coefficients, corresponding to p-type conduction, whereas positive chemical potentials ($\mu > 0$) produce negative values characteristic of n-type transport [103]. Distinct peaks are observed near the band edges, with |S| exceeding ±1000 μV/K at lower temperatures. As temperature increases, peak magnitudes decrease due to Fermi-Dirac broadening and bipolar transport effects [104,105].

As shown in Fig. 8(a), all compounds exhibit negative Seebeck coefficients over the 300-1000 K temperature range, which confirms the n-type behavior and electrons being the dominant charge carriers. With the increase of temperature, the absolute magnitude of |S| gradually decreases due to the increased thermal excitation of carriers across the band gap, which in turn reduces the Seebeck voltage. Among the studied compounds, the smallest absolute Seebeck coefficients are observed for $K_2SnGeI_6$ and $Rb_2SnGeI_6$, suggesting that the iodide-based compounds have a higher intrinsic carrier concentration compared to the chloride and bromide-based compounds.

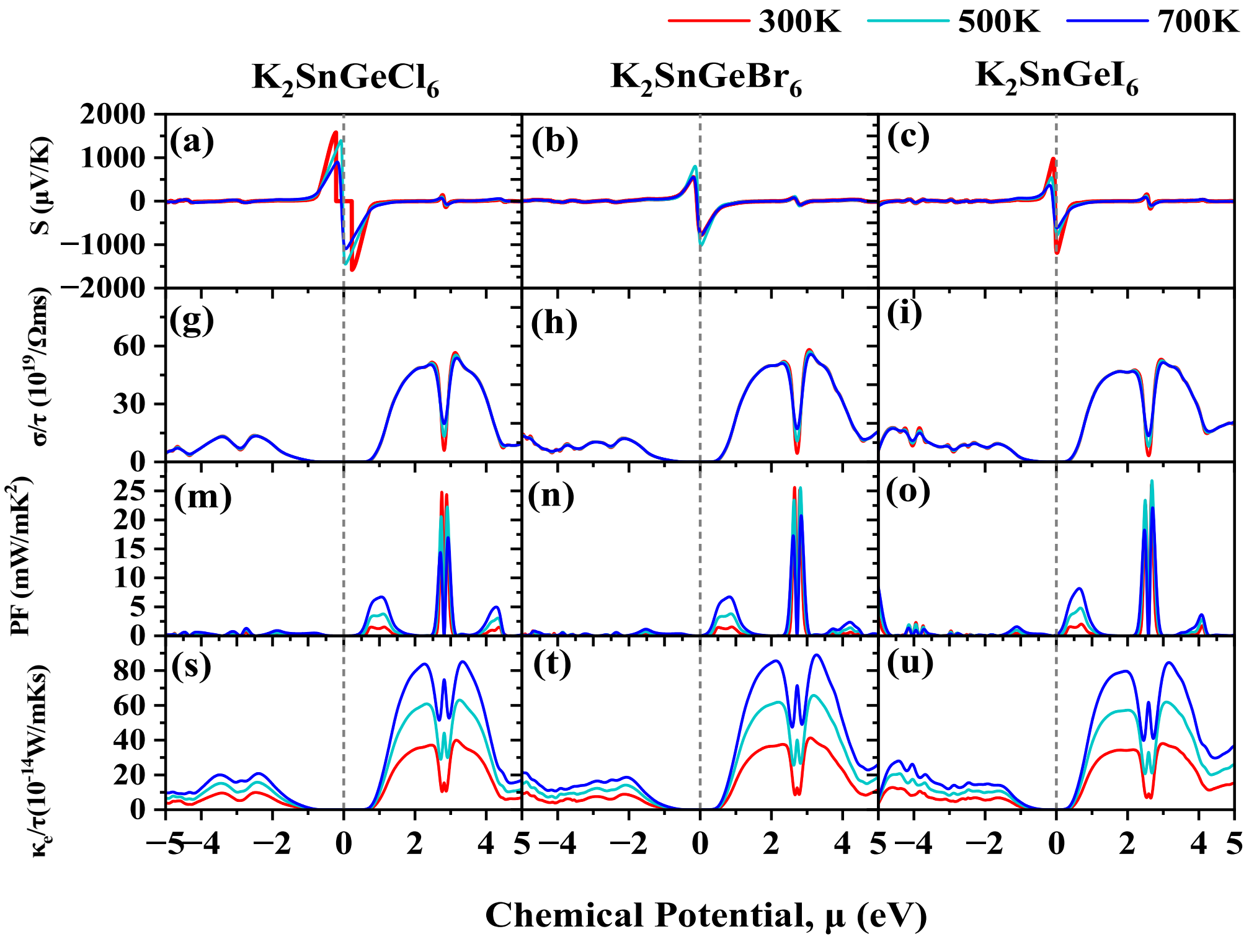

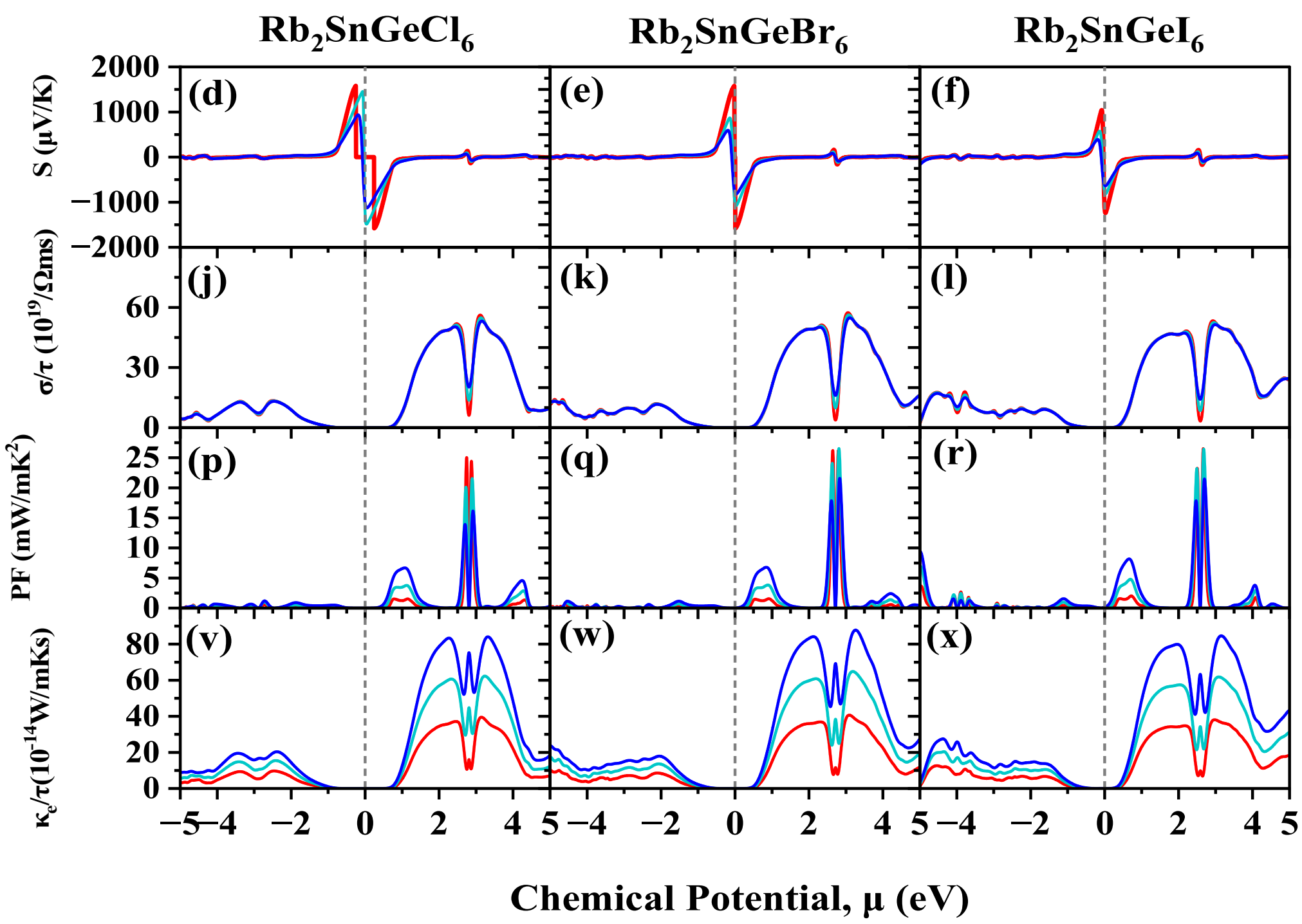

**Fig. 7.** Transport properties of $\beta_2SnGeX_6$ (β = K, Rb; X = Cl, Br, I) as a function of chemical potential and temperature.

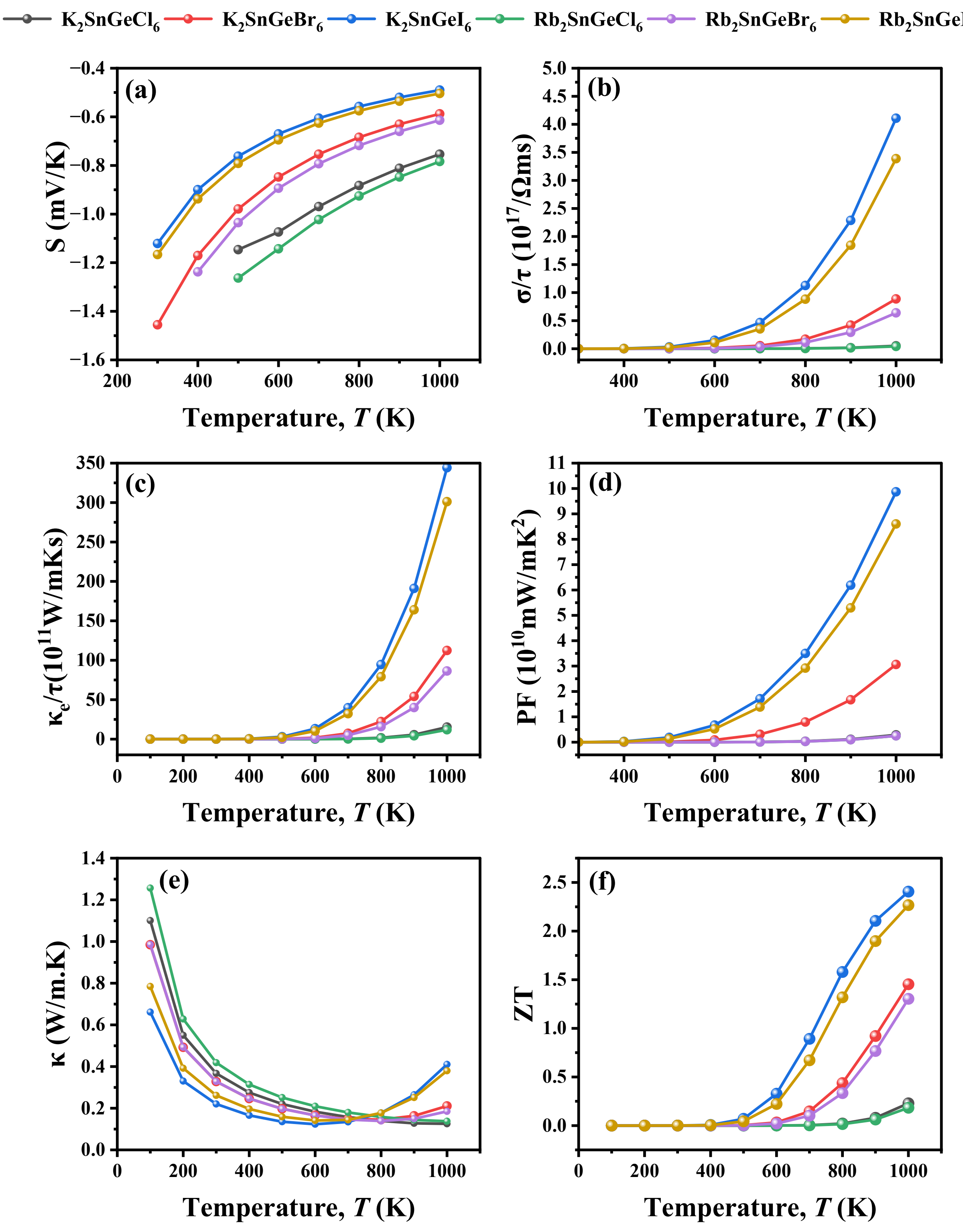

**Fig. 8.** Calculated transport properties of $\beta_2SnGeX_6$ (β = K, Rb; X = Cl, Br, I*)* as a function of temperature at constant chemical potential.

Electrical conductivity ($\sigma/\tau$) originates from the transport of charge carriers under an applied electric field, density gradioent or thermal gradient. Within the constant relaxation time approximation, $\sigma/\tau$ was evaluated as a function of chemical potential at various temperatures in **Fig. 7(g-l)** and as a function of temperature at a fixed chemical potential **Fig. 8(b).** In **Fig. 7(g-l)**, within the intrinsic region (−1.5 to 1 eV), $\sigma/\tau$ remains relatively low due to the limited carrier concentration. As the chemical potential approaches the valence or conduction band edges, $\sigma/\tau$ increases remarkably due to the enhanced participation of charge carriers in transport. While chemical potential is the dominant factor influencing $\sigma/\tau$, the temperature dependence is comparatively modest. Small reductions in $\sigma/\tau$ at specific energies correspond to features in the electronic band structure where the density of states is reduced [106,107].

In **Fig. 8(b)**, $\sigma/\tau$ increases almost exponentially with temperature for all studied compounds, which is consistent with thermal activation of carriers into the conduction band [108,109]. The iodide-based perovskites, particularly $K_2SnGeI_6$ and $Rb_2SnGeI_6$, show higher conductivity at elevated temperatures. This behavior can be explained by the lower electronegativity of iodine and its stronger orbital overlap, which favors the delocalization of the carriers and their enhanced mobility.

The electronic thermal conductivity shown in **Fig. 7(s-x)** and **Fig. 8(c)** $\kappa_e/\tau$ follows a trend analogous to $\sigma/\tau$, in agreement with the Wiedemann-Franz relation. Generally, $\kappa_e/\tau$ increases with both temperature and chemical potential due to the increased population of thermally activated carriers. In strict accordance with the Wiedemann-Franz law, electronic heat transport scales proportionally with electrical conductivity. Consequently, the highly conductive iodide-based compounds also exhibit enhanced electronic heat transport.

The power factor ($PF = S^2\sigma/\tau$) is a key parameter that determines the efficiency of thermoelectric materials, as it combines the contributions of the Seebeck coefficient and electrical conductivity. **Fig. 7(m-r)** presents the calculated PF as a function of chemical potential at different temperatures. As commonly observed in semiconductors, an increase in chemical potential leads to enhanced electrical conductivity but a reduced Seebeck coefficient. Consequently, the PF exhibits distinct peaks due to the competing behavior between S and $\sigma$. The magnitude of the PF peaks increases with temperature, indicating improved carrier transport and enhanced thermoelectric performance at elevated temperatures [110,111]. This trend suggests that $K_2SnGeX_6$ and $Rb_2SnGeX_6$ compounds may exhibit better thermoelectric efficiency under high-temperature operating conditions **Fig. 8(d)** reveals that the iodide compounds achieve the highest overall power factor (PF). Despite their relatively lower absolute Seebeck coefficients, the pronounced enhancement in electrical conductivity outweighs this decrease, resulting in higher PF values. In particular, $(K/Rb)_2SnGeI_6$ stands out as the best-performing system, highlighting iodide substitution as a highly effective approach for improving electrical transport behavior.

**Fig. 8(f)** presents the temperature dependence of the thermoelectric figure of merit ZT. At temperatures below 400 K, all materials exhibit low ZT values due to limited electrical conductivity. Above 500 K, the iodide-based compounds ($K_2SnGeI_6$ and $Rb_2SnGeI_6$) show a pronounced increase, reaching peak ZT values of approximately 2.4 and 2.25 at 1000 K, respectively. The high ZT values mainly arise from their high-power factor and low thermal conductivity. These values exceed the typical benchmark for high-performance thermoelectric materials, indicating that the iodide compounds are highly promising candidate for high-temperature waste heat recovery. In comparison, chloride-based compounds and $K_2SnGeBr_6$ remain low ($ZT < 0.3$), while $Rb_2SnGeBr_6$ exhibits intermediate performance, reaching around 1.3 at 1000 K.

# 4 Conclusion

In summary, this first-principles study systematically investigates the structural, mechanical, electronic, and transport characteristics of the novel lead-free halide double perovskite series $\beta_2SnGeX_6$ ($\beta$ = K, Rb; X = Cl, Br, I). Ground-state structural optimizations and geometric stability indices confirm that all six variations exhibit robust thermodynamic and structural stability within the highly symmetric cubic *Fm*-3*m* (#225) space group. Elastic constants and derived mechanical criteria characterize the entire series as fundamentally ductile, implying superior mechanical workability and resistance to micro-crack propagation during thin-film deposition and device processing. The high machinability index and dry lubricity of the compounds studied indicate that these are suitable for certain engineering applications [**112**-**115**]. Electronic band structure calculations reveal direct bandgaps located precisely at the high-symmetry Γ-point, demonstrating exceptional composition-dependent tunability from 1.44 eV down to 0.64 eV via progressive halogen substitution (Cl>Br>I).

The highly adaptable physical features yield multi-dimensional application profiles across several green-energy frontiers. The wide gap chloride variations ($K_2SnGeCl_6$ and $Rb_2SnGeCl_6$) possess optical bandgaps perfectly aligned with the Shockley-Queisser optimum for high efficiency single-junction photovoltaic absorbers, whereas the narrower gap bromide and iodide analogs are excellently suited for tandem solar sub-cells and near-infrared (NIR) photodetectors. Concurrently, their high static refractive indices and low zero-frequency reflectivity optimize internal light-trapping mechanisms, positioning these compounds as viable candidates for integrated anti-reflection coatings and optical waveguides. Beyond optoelectronics, the series displays outstanding capabilities for solid-state thermoelectric power generation. The incorporation of heavy constituent elements induces strong lattice anharmonicity and intense high-temperature Umklapp phonon-phonon scattering, which sharply depresses lattice thermal conductivity. When paired with low carrier effective masses that facilitate rapid charge transport and maximize electrical conductivity, the iodide variations achieve superior power factors and remarkable dimensionless figures of merit ($ZT \approx 2.4$ for $K_2SnGeI_6$ at 1000 K). Ultimately, by successfully eliminating toxic lead while preserving excellent optical, mechanical, and transport performance, the $\beta_2SnGeX_6$ family emerges as an environmentally benign, highly efficient, and versatile platform for next-generation sustainable electronics and high-temperature waste-heat recovery.

## Data availability

The data that support the findings of this study are available from the corresponding author upon reasonable request.

## Declaration of interest

The authors declare that they have no known competing financial interests or personal relationships that could have appeared to influence the work reported in this paper.

## CRediT authorship contribution statement

**Jubair Hossan Abir:** Conceptualization, Software, Methodology, Formal analysis, Data curation, Visualization, Writing-original draft, review & editing; **Tauhidur Rahman:** Software, Methodology, Formal analysis, Data curation, Visualization, Writing- draft, review & editing**;** **Sarker Sukanta Babu Pallab:** Formal analysis, Data curation, Visualization, Writing- draft, review & editing; **Md. Sharear Aman:** Writing- draft, review & editing; **Raihana Shams Islam:** Conceptualization, Supervision, Validation, Administration, Writing- Reviewing and Editing; **Saleh Hasan Naqib:** Conceptualization, Supervision, Validation, Administration, Writing-Reviewing and Editing.